\documentclass[10pt,aps,prb,twocolumn, nofootinbib]{revtex4-1}

\usepackage{amssymb,amsmath}
\usepackage{graphicx}
\usepackage{color}
\usepackage{hyperref}
\usepackage{cleveref}
\usepackage{listings}
\usepackage{natbib}
\usepackage{upgreek}

\newcommand{\p}{\partial}
\newcommand{\ep}{\varepsilon}

\newcommand{\nn}{\nonumber}
\newcommand{\ta}{\theta}

\newcommand{\wh}{\widehat}
\newcommand{\be}{\begin{equation}}                                             
\newcommand{\ee}{\end{equation}}
\newcommand{\ba}{\begin{eqnarray}}
\newcommand{\ea}{\end{eqnarray}}

\begin{document}

\title{\Large Frequency comb generation via cascaded second-order nonlinearities in microresonators}
\author{Jan Szabados$^{1}$, Danila~N.~Puzyrev$^{2}$, Yannick Minet$^{1,3}$, Luis Reis$^{1}$, Karsten~Buse$^{1,4,*}$, 
	Alberto Villois$^{2}$, Dmitry~V.~Skryabin$^{2,5,\dagger}$, and 
	Ingo~Breunig$^{1,4}$}
\affiliation{\footnotesize{
\mbox{$^{1}$Laboratory for Optical Systems, Department of Microsystems Engineering - IMTEK, University of Freiburg, Georges-K\"ohler-Allee 102,}\\ 79110 Freiburg, Germany\\
\mbox{$^{2}$Department of Physics, University of Bath, Bath BA2 7AY, United Kingdom}\\
\mbox{$^{3}$Gisela and Erwin Sick Chair of Micro-optics, Department of Microsystems Engineering - IMTEK, University of Freiburg,}\\  79110 Freiburg, Germany\\
\mbox{$^{4}$Fraunhofer Institute for Physical Measurement Techniques IPM, Heidenhofstra\ss e 8, 79110 Freiburg, Germany}\\
\mbox{$^{5}$Russian Quantum Centre, Skolkovo 143025, Russia}
}
\\$^*$karsten.buse@ipm.fraunhofer.de\\
$\dagger$d.v.skryabin@bath.ac.uk}

\begin{abstract}
Optical frequency combs are revolutionising modern time and frequency metrology. In the past years, their range of applications has increased substantially, driven by their miniaturisation through microresonator-based solutions. The combs in such devices are typically generated using the third-order $\chi^{(3)}$-nonlinearity of the resonator material. An alternative approach is making use of second-order $\chi^{(2)}$-nonlinearities. While the idea of generating combs this way has been around for almost two decades, so far only few demonstrations are known, based either on bulky bow-tie cavities or on relatively low-$Q$ waveguide resonators. Here, we present the first such comb that is based on a millimetre-sized microresonator made of lithium niobate, that allows for cascaded second-order nonlinearities. This proof-of-concept device comes already with pump thresholds as small as 2~mW, generating repetition-rate-locked combs around 1064~nm and 532~nm. From the nonlinear dynamics point of view, the observed combs correspond to the Turing roll patterns.
\end{abstract}

\date{\today}

\maketitle
\section{Introduction}
Optical frequency combs have been shown to be useful for applications such as precision spectroscopy in fundamental science,\cite{Newbury2011, Diddams2001} ultrafast dual-comb broadband single-pixel spectroscopy,\cite{Picque2019} ultrafast distance measurements\cite{Gaeta2019} and Tbit/s telecommunication,\cite{Gaeta2019} just to name some examples. Furthermore, they provide a framework for quantum signal and information processing.\cite{Kues2019} The range of applications has increased especially since the demonstration of comb generation in an ultra-high-$Q$ microresonator in 2007,\cite{Del'Haye2007} paving the way for the miniaturisation of these devices. Using  third-order $\chi^{(3)}$ (Kerr) optical nonlinearities has been the mainstream of research on microresonator frequency combs in recent years.\cite{Gaeta2019, Chembo2010, Kippenberg2011, Hansson2013, Kippenberg2018} Solitonic Kerr combs typically require anomalous dispersion at the pump frequency that is not readily available across the practically valuable visible part of the spectrum. They also require  an additional nonlinear element for $f$-$2f$-interferometry in order to stabilise the carrier envelope frequency.\cite{Gaeta2019} An obvious solution to this is to use a separate cavity with an intrinsic $\chi^{(2)}$-nonlinearity to convert the initial comb via second-harmonic and sum-frequency-generation.\cite{Herr2018} One approach to simplify this is to use materials for Kerr-comb generation coming with an intrinsic $\chi^{(2)}$-nonlinearity or to induce a $\chi^{(2)}$-nonlinearity by e.g., stress through material growth,\cite{Levy2011} ion migration\cite{Billat2017} or electric fields.\cite{Timurdogan2017} This way, one can generate a Kerr-comb and convert it in a single cavity as has been demonstrated in the $\chi^{(2)}$-materials aluminium nitride,\cite{Guo2018} gallium phosphide\cite{Wilson2018} and lithium niobate.\cite{He2019} There is, however, a more elegant way to generate frequency combs at both the pump and its second-harmonic simultaneously based on $\chi^{(2)}$-nonlinearities only. This approach relies on cascaded second-order nonlinear-optical processes, i.e.\,frequency doubling followed by optical parametric generation, again leading to frequency doubling and so on.\cite{Buryak2002} Since this process is intrinsically linked to comb generation around both the pump and second-harmonic frequencies,\cite{ol1} this approach is automatically suitable for self-referencing. Furthermore, it promises to provide much lower pump thresholds since for continuous-wave light the second-order nonlinearity is generally much stronger than the third-order one. Although the idea has been around for almost two decades, so far only few experimental realisations are known, based either on bulky bow-tie cavities\cite{Ulvila2013, Ulvila2014, Ricciardi2015, Mosca2016, Mosca2018} or on relatively low-$Q$ waveguide resonators,\cite{Ikuta2018} but none in high-$Q$ microresonators. 
\begin{figure*}[]
	\centering
	\includegraphics[width=0.9\textwidth]{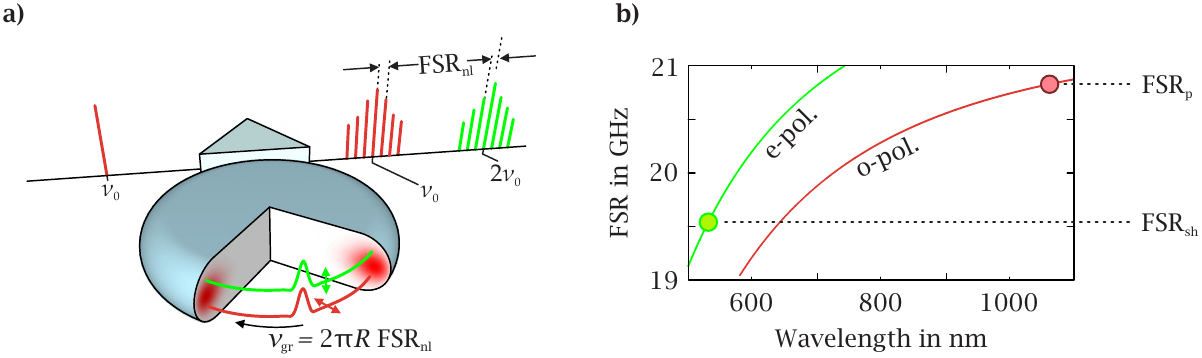}
	\caption{\small \textbf{Microresonator and simultaneous near-infrared and visible frequency comb generation via $\chi^{(2)}$-nonlinearity.} \textbf{a)} Continuous-wave near-infrared laser light with frequency $\nu_0$ is prism-coupled into a microresonator (radius $R$). When the phase-matching-criterion for second-harmonic generation is fulfilled, second-harmonic light at  $2\nu_0$ is generated. A sequence of cascaded second-order nonlinear events then leads to the build-up of the interlocked near-infrared (red spectrum) and visible (green spectrum) combs. In the comb generation regime, the ordinarily (o) polarised pump and extraordinarily (e) polarised second-harmonic wave forms travel with the same group velocity $v_\mathrm{gr}$, associated with the repetition rate equaling the free spectral range (FSR) FSR$_\mathrm{nl}$. \textbf{b)} FSRs of the linear resonator vs wavelength:  FSR$_\mathrm{p}=20.8$~GHz and FSR$_\mathrm{sh}=19.5$~GHz are the FSRs for the pump and second-harmonic.}
	\label{fig1}
\end{figure*}
\section{Results}
Below we present the first frequency comb generated using cascaded second-order nonlinearities in a millimetre-sized ultrahigh-$Q$ ($Q>10^{8}$) whispering gallery resonator  made of 5\% MgO-doped z-cut congruent lithium niobate and pumped with continuous-wave near-infrared laser light. The narrow-band pump light is converted inside the resonator into two combs with side-bands growing around the pump itself and its second-harmonic (Fig.\,\ref{fig1}a).
If the resonator is pumped with ordinarily (o) polarised light and the birefringence-phase-matching condition for second-harmonic generation is fulfilled, then extraordinarily (e) polarised second-harmonic light is generated.\cite{Fuerst2010} More details on the sample fabrication and experimental arrangements are provided in Supplementary Information. For pump powers exceeding a threshold the primary sidebands are generated around the pump and signal wavelengths. Subsequently, the side-band generation process develops in a cascaded fashion, building up frequency comb structures around both the pump and the second-harmonic frequencies. Some prior theory and modelling of these processes in second-harmonic and down-conversion arrangements can be found in literature.\cite{ol1,Hansson2016,ol2,opex1} For the microresonator used, the free spectral ranges (FSRs) of the linear cavity at the pump and second-harmonic frequencies are FSR$_\mathrm{p}=20.8$~GHz and FSR$_\mathrm{sh}=19.5$~GHz (Fig.\,\ref{fig1}b). However, as we will demonstrate below, the FSRs of the combs around the pump and second-harmonic become locked to the same nonlinear FSR, FSR$_\mathrm{nl}$, corresponding to a shared repetition rate (group velocity) of the wave forms emerging at both frequencies (Fig.\,\ref{fig1}a).
\begin{figure*}[]
	\centering
	\includegraphics[width=0.9\textwidth]{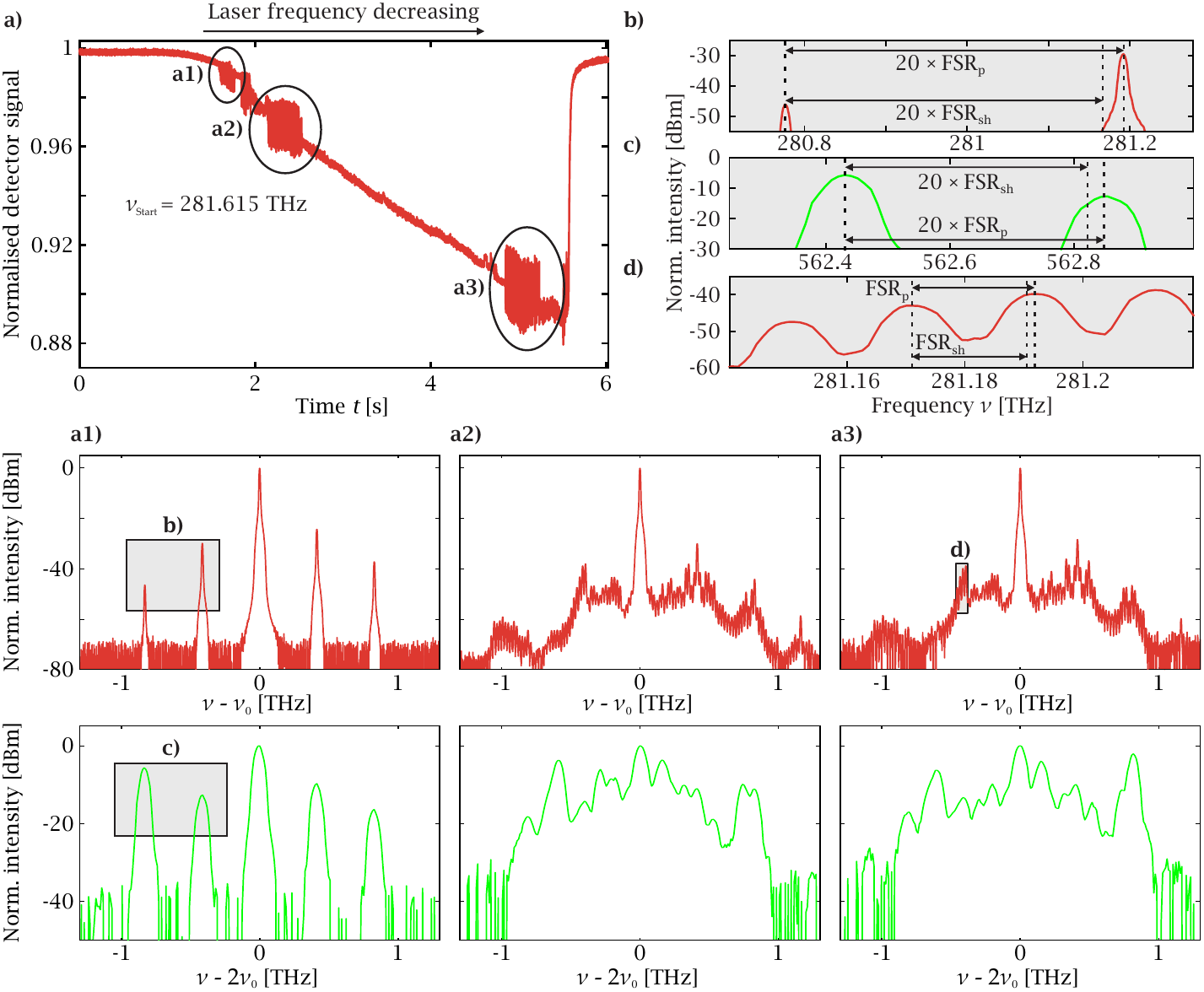}
	\caption{\small \textbf{Experimental frequency comb signals.} \textbf{a)} Slow scan of the pump laser frequency $\Omega$ across the pump mode used for comb generation. Reducing $\Omega=2\pi\nu_0$ reveals frequency comb generation areas marked a1), a2) and a3). At a1), we observe four sidebands separated by $\simeq 20\times$FSR$_{p}$ around the pump and second-harmonic  frequencies. Bringing $\Omega$ into the a2) range reveals dense quasi-continua of modes again simultaneously generated around the pump and second-harmonic frequencies. With further reduction of the frequency the system enters a quasi-stationary state, which is replaced again by the dense quasi-continua in the a3) range. Subsequently the resonance is lost. \textbf{b),c),d)} A close-up revealing that the individual peaks of the sparse spectra are separated by $\simeq 20\times$FSR$_{p}$ at both the pump, see b), and the second harmonic, see c). As modelling shows the actual nonlinear FSR is slightly, by $\sim 1$MHz, less than FSR$_p$.	Quasi-continua spectra at the pump frequency naturally reveal the separation of the neighboring lines  by $\simeq$FSR$_{p}$, see d), while at the  second-harmonic wavelength the optical spectrum analyser does not allow to resolve features on the FSR-scale.}
	\label{fig2}
\end{figure*}  
The total quality factor at the pump frequency is determined to be $Q=2.8\times10^{8}$ in the coupling regime used.\cite{Breunig2016} The maximum coupling efficiency\cite{Breunig2016} we achieve is $K_\mathrm{max}=0{.}18$, limited by imperfect spatial mode overlap only. Thus, with the laser pump power being 10.3~mW, the maximum incoupled pump power is approximately $K_\mathrm{max}\times 10{.}3~\mathrm{mW}\approx 1{.}9$~mW. When the laser frequency is reduced,  the pump resonance is broadened through the interplay of thermal and nonlinear effects (Fig.\,\ref{fig2}a)) and effectively stretches the frequency detunings relative to the cold cavity reference.\cite{Ilchenko1992} We observe multiple frequency comb states, marked a1), a2) and a3) in Fig.\,\ref{fig2} while performing a scan. At a1), only four spectral sidebands can be observed at both the pump and second-harmonic wavelengths. The spacing of these sidebands at both the pump and second-harmonic is approximately 416~GHz, which corresponds to 20~FSRs of the linear resonator at the pump wavelength, FSR$_\mathrm{p}$. Thus, the FSR of the comb spectrum even at the second-harmonic locks to a value close to the linear pump FSR, i.e., FSR$_\mathrm{nl}\simeq\mathrm{FSR}_\mathrm{p}$. This is striking, considering the large FSR-offset of $1.3$~GHz in the linear regime (Figs.\,\ref{fig2}b-\ref{fig2}d)), that would accumulate to the noticeable $\simeq 26$~GHz across $20$ modes of the linear resonator.  
This is, we believe, the first demonstration of an FSR-locked fundamental-to-second-harmonic frequency comb generated solely due to  $\chi^{(2)}$-nonlinearity in a microresonator. 
The next frequency range, a2), being around the smaller laser frequency $\Omega=2\pi\nu_0$, corresponds to much broader, $2$-THz wide, near-infrared and green spectra (Fig.\,\ref{fig2}a2)). Similar spectra are also observed in the a3) range. A close-up of Fig.\,\ref{fig2}a3) (top) reveals the individual peaks to be  indeed separated by  $\simeq 20.8$~GHz, which is very close to the linear FSR$_\mathrm{p}$ at the pump wavelength. Thus, the obtained comb consists of about $100$ individual comb lines. We have also observed a broad spectrum around the second-harmonic frequency, see  Fig. \ref{fig2}a3) (bottom). However, in this case, due to the limited resolution of the optical spectrum analyser employed, the individual comb lines cannot be resolved and the FSR locking is confirmed only in numerical simulations, see  Figs.\,\ref{fig4}c) and \ref{fig4}f). 
\begin{figure}[]
	\centering
	\includegraphics[width=0.49\textwidth]{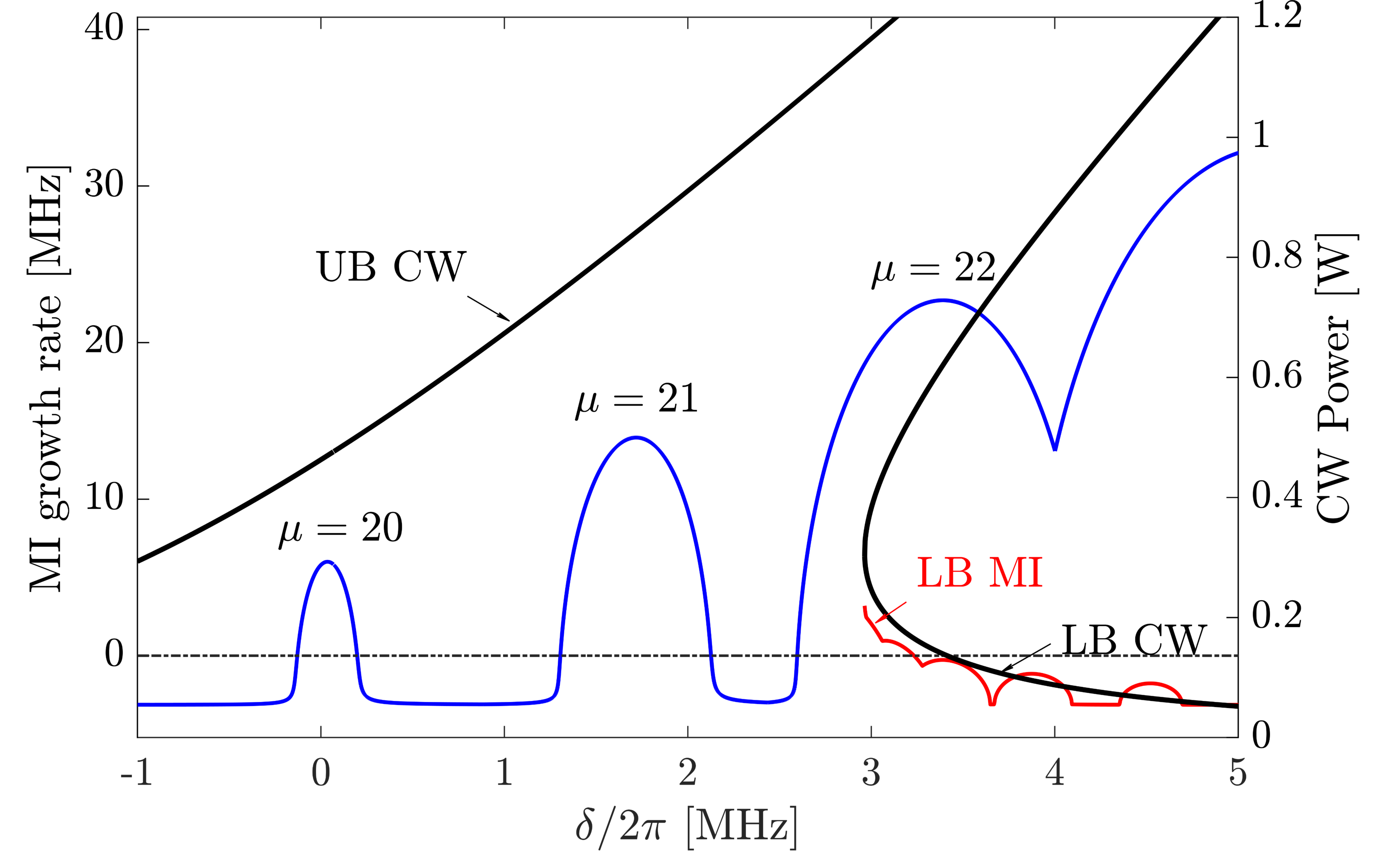}
	\caption{\small \textbf{Modulational instability (MI) growth rate.} Blue line shows the MI growth rate of the upper branch (UB) of the intracavity cw-state vs the cold-cavity detuning $\delta=\omega_\mathrm{p}-\Omega$. $\mu=\pm 20, \pm 21, \pm 22$ are the mode number offsets with the highest gain within a given interval of $\delta$'s. $\mu$'s are counted left and right from the mode number, $m_p=13112$, corresponding to the cavity resonance $\omega_\mathrm{p}$ nearest to the pump frequency $\Omega$. $\mu=\pm 20$ matches the experimentally observed spectral lines separated by $\simeq 20$ FSR. Bold black lines and the right vertical axis show the peak power of the intracavity pump cw state vs $\delta$. The bistability interval starts from $\delta/(2\pi)\simeq 3$~MHz and extends towards larger $\delta$'s. Red line shows MI growth rate of the lower branch (LB) of the  bistability loop. The dashed horizontal line indicates zero of the MI gain. One can see that LB is stable through most of its existence range.
	}
	\label{fig3}
\end{figure} 
We proceed now with introducing a more theoretical insight into the mechanisms leading to the frequency comb generation in our system. Spectral data in Fig.\,\ref{fig2}a1) suggest that we are dealing with modulation-instability (MI) induced spectral broadening. The MI in our context is interpreted as the generation  of selected resonator modes detuned away from the pump. 
In particular, through the MI analysis (see Methods) we have estimated the sideband mode numbers 
$M=m_\mathrm{p,s}\pm \mu$ that acquire the maximal growth rates. Here $m_\mathrm{p}$ is a mode number corresponding to the cold resonator frequency $\omega_\mathrm{p}$ that is nearest to the laser frequency $\Omega$, $m_\mathrm{s}=2m_\mathrm{p}$ and $\mu$ is the mode number offset.
When sidebands grow appreciably, then they start serving as effective pumps themselves triggering a cascade of up- and down-conversion events between the near-infrared and green spectra. This is the mechanism leading to frequency comb generation in our setting.
Figure~\ref{fig3} shows the calculated MI growth rate as a function  of the cold resonator detuning $\delta=\omega_\mathrm{p}-\Omega$ with the maximally unstable $\mu$ being  $\pm 20,\pm 21,\pm 22$. 
In particular, $\mu=20$  corresponds to the first spectra measured during  the  experimental scan.
This state is characterised by line separation of $\simeq 20\times$FSR$_\mathrm{nl}$ (Fig.\,\ref{fig2}a1)). 
For larger  $\delta$ (smaller $\Omega$) the MI gain disappears and then reappears again with $\mu=21$ becoming  most unstable. Most unstable modal indices continue to alternate with further increase of $\delta$ until the resonance is lost (Fig.\,\ref{fig3}). While MI analysis explains how combs are triggered, dynamical modelling of the full nonlinear system is required to check what spectral shapes are generated after many resonator round-trips. 
Within the MI intervals of $\delta$ we numerically generate spectra qualitatively and quantitatively similar to the experimental ones and in the same sequential order (Fig.\,\ref{fig4}). In the $\mu=20$ interval centred at $\delta\approx 0$, the near-infrared and green comb spectra have four sparse sidebands separated by $\simeq 20\times$FSR$_\mathrm{nl}$. In the $\mu=21$ and $\mu=22$  intervals, the comb spectra become quasi-continuous and extend over $\simeq 2$~THz bandwidth. For $\delta\gtrsim 3$~MHz the comb spectra are not observed.  
We note, that alternating stable and unstable detuning intervals (cf.\,experimental Fig.\,\ref{fig2}a and numerical Fig.\,\ref{fig4}a) with discretely varying $\mu$'s  unambiguously indicate that the field dynamics in a microresonator is strongly influenced by the relative sparsity of its spectrum. While previous studies of MI inside more dense spectra, such as in large bow-tie $\chi^{(2)}$ cavities, did not, and likely, could not, reveal the above features.\cite{Mosca2018}
\section{Discussion}
\begin{figure*}[]
	\centering
	\includegraphics[width=0.9\textwidth]{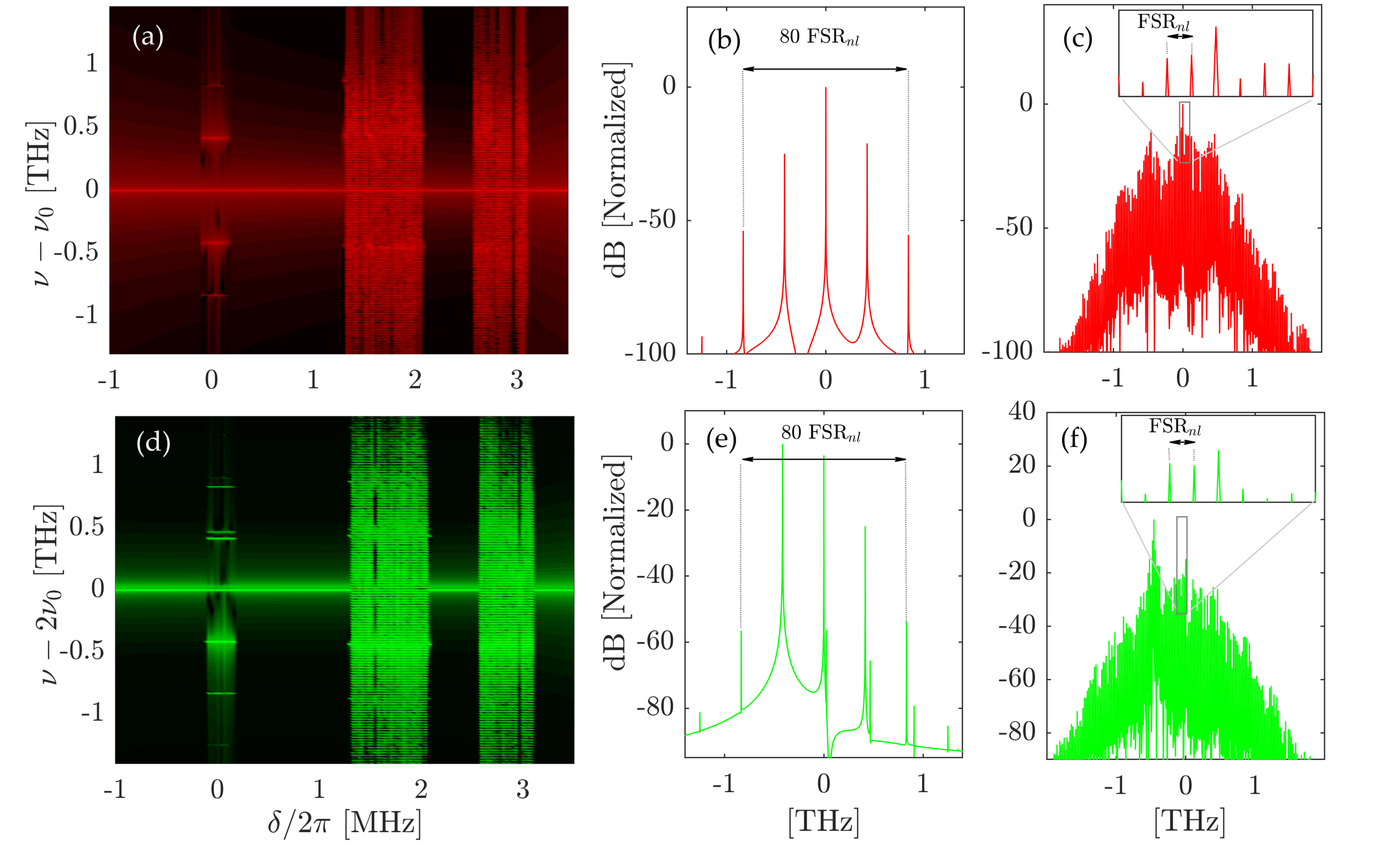}
	\caption{\small \textbf{Spectra observed in numerical simulations across the detuning range as in Fig.\,\ref{fig3}.}
		Top/bottom row is for the pump (red)/second-harmonic (green) field. \textbf{a,d)} Numerically calculated spectra for a varying detuning $\delta$ showing three intervals of the comb generation. \textbf{b,e)} show the near-infrared and green spectra for $\delta/2\pi=0.0699$~MHz from a) and d), respectively. \textbf{c,f)} show the spectra for $\delta/2\pi=2.6189$~MHz. Zoomed areas show the individual resonances in the quasi-continua separated by the FSR$_\mathrm{nl}$ for both the near-infrared and green spectra. To achieve the spectral states shown here the total number of resonator round trips used was 208,000. 
	}
	\label{fig4}
\end{figure*}

The sparse spectra (Figs.\,\ref{fig2}a1), \ref{fig4}b), \ref{fig4}c)) correspond to weakly-modulated nonlinear periodic wave trains in the real space, that are associated with the so-called Turing roll patterns (Supplementary Figs.\,1-3). These type of patterns have been extensively studied recently in Kerr microresonators.\cite{tur1,tur2,tur3} An important and distinct feature of the $\chi^{(2)}$ Turing patterns reported by us is that they correspond to a two-colour signal with components that are inherently connected nonlinearly and at the same time spectrally separated by an octave. The linear cavity repetition rate varies substantially from one spectral end to the other, as was already discussed above. However, nonlinear effects are able to compensate for this difference. Indeed, we can resolve numerically both sparse and quasi-continuous spectra and find the same FSR$_\mathrm{nl}$ for the near-infrared and green spectra, such that FSR$_\mathrm{p}$-FSR$_\mathrm{nl}\simeq 1$~MHz. It means that the FSR is locked to a value shifted slightly away from the FSR$_\mathrm{p}$ and very significantly, by $\simeq 1.3$~GHz, from the FSR$_\mathrm{sh}$. This asymmetry is largely due to the fact that the power of the pump is much stronger than the signal.  
The real space wave trains associated with the quasi-continua in the $\mu=21,~22$ intervals emerge through the destabilisation of Turing patterns and have much deeper modulations of the field intensity (Supplementary Fig.\,2). Despite this, the locking of FSRs to FSR$_\mathrm{nl}$ also appear to persist for the quasi-continuous spectra. If $\delta$-values are kept outside the bistability of the homogeneous intracavity state, $\delta/2\pi\lesssim 3$~MHz (Fig.\,\ref{fig3}), then the continua persist indefinitely, since they have no alternative state to transform into. However, as soon as the system becomes  bistable, a stable low-amplitude state serves as a strong attractor that quickly drags the system outside the continuum and into a low and constant intensity stable operation regime. The latter corresponds to the abrupt loss of the resonance in the experimental measurements and modelling. \\
FSR locking across the octave may appear to be counterintuitive at first glance. However, it is to be expected for weakly modulated Turing patterns. Indeed, for a constant amplitude solution the repetition rates, i.e.\,FSRs, lose their relevance. 
Hence, for weakly modulated (around such a solution) wave trains, the nonlinearity and dissipation induced locking effects should allow to shift FSR$_\mathrm{sh}$ to the one of the pump value. While, the FSR locking mechanisms for Turing patterns require further detailed studies, this  effect for the short solitonic pulses in an intracavity OPO has been studied by one of us two decades ago.\cite{skr1} Then, it was demonstrated only for the angular velocity mismatches comparable to the nonlinear linewidth  of the cold cavity ($\sim$ few to $10$s of MHz here),\cite{skr1} which is much less than the  $\sim 1$~GHz velocity mismatches relevant for us here.  
The results presented here can be considered a first step towards unveiling the full potential of microresonator-based frequency combs originating from cascaded second-order nonlinear-optical processes. While we do not show a soliton comb in this contribution, theoretical studies have argued that it is possible to generate such a comb with the scheme used providing the phase matching is accompanied by a reduced or vanishing FSR-offset between the pump and second-harmonic.\cite{skr1,opex1,ol1} 
FSR matching points spectrally separated by an octave can be found in the present sample, but then the phase matching must be provided through quasi-phase-matching (QPM) technique.\cite{Breunig2016} Introducing a QPM-structure would furthermore allow a great degree of flexibility to choose the centre wavelengths and polarisation states of the generated combs. While there is still a lot of room for improvement, this new way of generating frequency combs comes with the advantage of using the generally stronger $\chi^{(2)}$-nonlinearities as opposed to the $\chi^{(3)}$-nonlinearities used in microresonator-based combs so far. We already observe combs at pump thresholds as low as $2$~mW. 
Highly-optimised Kerr comb systems with thresholds of slightly below $1$~mW have only very recently been demonstrated, while in resonators of similar size to the one used here the best thresholds were approximately $3$~mW.\cite{Zhang2019} Thus, we expect the proof-of-principle device demonstrated here to initiate significant research activities.\\
Methods used in this contribution can be applied to any material exhibiting $\chi^{(2)}$-nonlinearities. Thus, one could think of trying to access more challenging wavelength regimes such as the mid-infrared using non-oxide materials like AgGaSe$_{2}$, for which the essential step of OPO in a microresonator has already been demonstrated.\cite{Meisenheimer2017} While we rely on hand-polished microresonators here, the method introduced can also be employed for batch-compatible chip-integrated microresonators. Recently,  the first OPO has been demonstrated on a chip-integrated platform, potentially paving the way for chip-integrated comb sources.\cite{Bruch2019} This would increase the scalability of the platform introduced here, making it even more appealing for applications.

\bibliographystyle{unsrt}

\begin{acknowledgments}
	This work was supported by the Fraunhofer and Max Planck Cooperation Program COSPA and by the Horizon 2020 Framework Programme (812818, MICROCOMB). Y.M. appreciates the support by a Gisela and Erwin Sick Fellowship. D.V.S. and A.V.
	are supported by the Leverhulme Trust (RPG-2015-456). The authors would like to thank H. Giessen (University of Stuttgart) for support regarding experimental equipment.
\end{acknowledgments}

\section*{Author contributions}
	J.S. conceived, designed and set up the experiment, analysed the data and prepared the figures. Y.M. and L.R. manufactured the microresonator and helped set up the experiment. D.V.S. conceived the governing equations. D.N.P., A.V. and D.V.S. designed codes, performed numerical modelling, analysed numerical and experimental data and prepared the figures.  D.N.P. found the regime with MI gain oscillating with cavity detuning leading to a good match between numerical and experimental data. J.S., K.B., I.B. and D.V.S. wrote the manuscript. K.B., I.B. and D.V.S. supervised the project.
	
\section*{Methods}

\textbf{Theory and modelling.} Our numerical modelling is based on an approach where the envelopes of the fundamental, $\psi_\mathrm{p}$, and second harmonic, $\psi_\mathrm{s}$, fields are distributed in the angular variable $\theta\in [0,2\pi)$ and evolve in time, $t$.
The corresponding model\cite{ol1} for $2m_\mathrm{p}=m_\mathrm{s}$, where $m_\mathrm{p,s}$ are the angular momenta of the birefringence matched fundamental and second harmonic, adopts notations for the dispersion coefficients $D_\mathrm{1p,1s,2p,2s}$ as introduced in literature:\cite{herr} 
\ba
&&\nn i\p_t\psi_\mathrm{p}=\left(\delta-iD_\mathrm{1p}\p_\ta-\tfrac{1}{2}D_\mathrm{2p}\p^2_\ta\right)\psi_\mathrm{p}-i\kappa_\mathrm{p}\psi_\mathrm{p}\\ && -\gamma_\mathrm{2p}\psi_\mathrm{s}\psi_\mathrm{p}^*e^{i\ep t}+\kappa_\mathrm{p}\sqrt{\eta FP},\label{e1}\\
&& \nn i\p_t\psi_\mathrm{s}=\left(2\delta-iD_\mathrm{1s}\p_\ta-\tfrac{1}{2}D_\mathrm{2s}\p^2_\ta\right)\psi_\mathrm{s}-i\kappa_\mathrm{s}\psi_\mathrm{s}\\  && -\gamma_\mathrm{2s}\psi_\mathrm{p}^2  e^{-i\ep t}.\label{e2}\ea
Here $\delta=\omega_\mathrm{p}-\Omega$ with $\omega_\mathrm{p}=2\pi\times 281.61$~THz being the cold cavity resonance frequency and $\Omega$ is the pump laser frequency. $\epsilon=2\omega_\mathrm{p}-\omega_\mathrm{s}$ is the cold cavity frequency mismatch parameter, which we allow to vary approximately within  $\pm 1.5$~FSRs. This is our only fitting parameter used to achieve a correspondence of the numerical and experimental results, see  more details below. $\omega_\mathrm{s}$ is the cold resonance frequency for the mode number $m_\mathrm{s}=2m_\mathrm{p}$. The physical fields $E_\mathrm{p,s}$ are reconstructed as $E_\mathrm{p}=\psi_\mathrm{p}(t,\ta)e^{-i\Omega t+im_\mathrm{p}\ta}+c.c.$, $E_\mathrm{s}=\psi_\mathrm{s}(t,\ta)e^{-i2\Omega t+im_\mathrm{s}\ta}+c.c.$ and are normalised to be measured in W$^{1/2}$.\cite{ol1} Time-domain spectra are calculated as Fourier transforms of $E_\mathrm{p,s}$ in $t$ for a fixed $\ta$. $D_\mathrm{1p}=2\pi\times 20.828$~GHz and $D_\mathrm{1s}=2\pi\times 19.519$~GHz are the FSR parameters for the two fields. 
$D_\mathrm{2p}=-2\pi\times 98$~kHz and $D_\mathrm{2s}=-2\pi\times 218$~kHz are the second-order dispersion parameters, corresponding to the normal dispersion at both frequencies. Dispersion has been calculated using COMSOL for the fundamental mode families in the ordinary and extra-ordinary polarisation states and checked against the asymptotic equations.\cite{book} $\kappa_\mathrm{p}=2\pi\times 0.5$~MHz and $\kappa_\mathrm{s}=2\pi\times 10$~MHz are the halves of the resonance widths, corresponding to the quality factors of $2.82\times 10^8$ and $2.82\times 10^7$. Nonlinear parameters $\gamma_\mathrm{2p}\simeq 1.2902\times 10^9$~s$^{-1}$W$^{-1/2}$, $\gamma_\mathrm{2s}\simeq 2.4956\times 10^9$~s$^{-1}$W$^{-1/2}$ are defined as in literature.\cite{ol1}
$P=1.854$~mW is the pump power. $\eta=\kappa_\mathrm{ex}/\kappa_\mathrm{p}\simeq 0.31$ is the coupling coefficient, where $\kappa_\mathrm{ex}$ is a fraction of the net loss ($\kappa_\mathrm{p}$) due to coupling of light in and out. Intracavity power is enhanced by the finesse $F=D_\mathrm{1p}/(2\kappa_\mathrm{p})\simeq 20000$. $\kappa_\mathrm{p}$ in front of the square root in the pump term provides the necessary scaling, so that in the linear regime and on the resonance $|\psi_\mathrm{p}|^2=\eta FP$. Eqs.\,(1,2) were solved numerically using a split-step approach with the 4th order Runge-Kutta method.

CW-states describing nonlinear reshaping of the cavity resonances and the associated bistability can be found using the substitutions  $\psi_\mathrm{p}(t,\ta)=\psi_{p0}$
and $\psi_\mathrm{s}(t,\ta)=\psi_\mathrm{s0}e^{-i\ep t}$, and solving the resulting algebraic equations numerically with a Newton method. 
To study MI we have added small perturbations $x_{1,3}$  to the field envelopes: $\psi_\mathrm{p}=\psi_\mathrm{p0}+x_1(\ta,t)$, $\psi_\mathrm{s}=(\psi_\mathrm{s0}+x_3(\ta,t))e^{-i\ep t}$. A system of equations for $x_{1,3}$ was linearised and extended by adding a pair of the complex conjugated equations for $x_1^*\equiv x_2$ and $x_3^*\equiv x_4$ leading to $i\p_t\vec x=\wh L(-i\p_\ta)\vec x$, where $\vec x=(x_1,x_2,x_3,x_4)^T$.
Here \ba
\nn &&\wh L(-i\p_\ta)=\left(\begin{array}{cccc}
	\wh A_\mathrm{p}&-\gamma_\mathrm{2p}\psi_\mathrm{s0}&-\gamma_\mathrm{2p}\psi_\mathrm{p0}^*&0\\
	\gamma_\mathrm{2p}\psi_\mathrm{s0}^*&-\wh A_\mathrm{p}^*&0&\gamma_\mathrm{2p}\psi_\mathrm{p0}\\
	-2\gamma_\mathrm{2s}\psi_\mathrm{p0}&0&\wh A_\mathrm{s}&0\\
	0&2\gamma_\mathrm{2s}\psi_\mathrm{p0}^*&0&-\wh A_\mathrm{s}^*
\end{array}\right),\\
\nn && \wh A_\mathrm{p}=\delta+D_\mathrm{1p}(-i\p_\ta)+\tfrac{1}{2}D_\mathrm{2p}(-i\p_\ta)^2-i\kappa_\mathrm{p},\\
\nn && \wh A_\mathrm{s}=2\delta-\ep+D_\mathrm{1s}(-i\p_\ta)+\tfrac{1}{2}D_\mathrm{2s}(-i\p_\ta)^2-i\kappa_\mathrm{s}.\ea
Assuming $\vec x(\ta,t)=\vec x_+ e^{i\mu\ta+\Lambda t}+\vec x_- e^{-i\mu\ta+\Lambda^* t}$ we solved numerically an eigenvalue problem for a $4\times 4$ matrix $\hat L(\mu)$ to find the MI growth rate $Re(\Lambda)>0$ vs  the mode number offset $\mu=0,\pm 1,\pm 2,\dots$ and other parameters. 
As $\ep$ is reduced from zero towards its negative values, then the most unstable $\mu$ that appears for the largest $\Omega$ (onset of the experimental scan) starts increasing in unitary steps with  $\mu=20$ 
being  achieved for $\ep=-1.17 D_\mathrm{1p}$. This $\ep$ was used to 
generate all the numerical data for this manuscript.

Supplementary Materials provide extra text and graphical data facilitating understanding of the device fabrication and characterisation, and of the modelling of the Turing patterns (Supplementary Figs.\,1-3). All numerical data used to produce this publication will be made available open access at https://researchdata.bath.ac.uk/ after publication of this work.
 ~\\
 ~\\
\newpage
\onecolumngrid
\appendix
\textbf{\Large \begin{center}
		Supplementary Information and Figures
	\end{center} }	
	\textbf{Microresonator fabrication.} We employ a femtosecond laser source emitting at a wavelength of 388~nm with a repetition rate of 2~kHz and 400~mW average output power to cut out a cylindrical preform from a 300-$\upmu$m-thick 5\% MgO-doped congruent lithium niobate (CLN) wafer. This material is chosen as it is readily available, low-cost and comes with strong $\chi^{(2)}$-nonlinearities, making it one of the most widely employed materials for $\chi^{(2)}$-nonlinear optics. Then, the cylinder is glued to a metal (brass) post for easier handling. Subsequently, we use the femtosecond laser source again to shape a whispering-gallery resonator with a major radius of $R=1$~mm and a minor radius of $r=380$~$\upmu$m as shown in Fig.\,\ref{fig4s}a). After this, to achieve optical-grade surface quality, we polish the resonator rim with a diamond slurry.
	
	\textbf{Additional experimental details.} The experimental setup is visualised in Fig.\,\ref{fig4s}b). The laser source used for our experiments is a commercially available NKT Koheras Basik Y10 fibre laser with a narrow linewidth of approx.\,$20$~kHz and a maximum output power of approx.\,10~mW. The laser is fibre-coupled with polarisation controllers attached to it. It is focused using a gradient-index lens to couple light polarised in the x-y-plane to the microresonator via a rutile prism. Since the resonator is made of z-cut CLN, this means that the incoupled light is polarised perpendicularly to the optic axis; thus, it experiences the ordinary refractive index.  To measure the linewidth of the cavity resonances, we calibrate the laser frequency change over time using a commercially available wavelength-meter (HighFinesse WS6-600). The power is kept very low at a few $\upmu$W to avoid thermal and nonlinear effects. The transmission spectrum is recorded with a silicon photodiode. This measurement allows to determine the quality factor according to the formula $Q=\nu_\mathrm{p}/\nu_\mathrm{FWHM}$, where $\nu_\mathrm{p}$ is the centre frequency of the cavity resonance and $\nu_\mathrm{FWHM}$ corresponds to the linewidth. The maximum coupling efficiency can be determined from the depth of the cavity resonance in critical coupling. To determine the quality factor in the coupling regime used, we use a model found in literature.\cite{Breunig2016}\\
	To monitor the obtained frequency spectra, we use a Yokogawa AQ6370D optical spectrum analyser for the near-infrared light around the pump frequency and a Yokogawa AQ6373B for the visible light around the second-harmonic frequency. The resonator is heated to temperatures $T\approx70$~$^{\circ}$C to fulfil the phase-matching-criterion to allow for efficient second-harmonic generation. This helps when designing the experimental setup: the o-polarised light around the pump and the e-polarised light around the second-harmonic are separated due to the large birefringence of the rutile prism used for coupling. To identify cavity resonances of interest, we scan the pump laser frequency over multiple GHz in order to observe sidebands around the pump wavelength. Once we have found a suitable resonance providing such sidebands, we tune the laser towards resonance by reducing its frequency to be able to make use of thermal locking.\cite{Carmon2004} The thermal locking process effectively stretches the resonance, so that the modelling that uses the cold cavity approximation reproduces experimental data for much smaller values of the detuning. To obtain the shape of the cavity resonance experimentally, we reduced the pump laser frequency by about 140~MHz in approximately 6~seconds.		
	\newpage
	\textbf{Supplementary figures.}

	\begin{figure*}[h]
		\centering
		\includegraphics[width=0.9\textwidth]{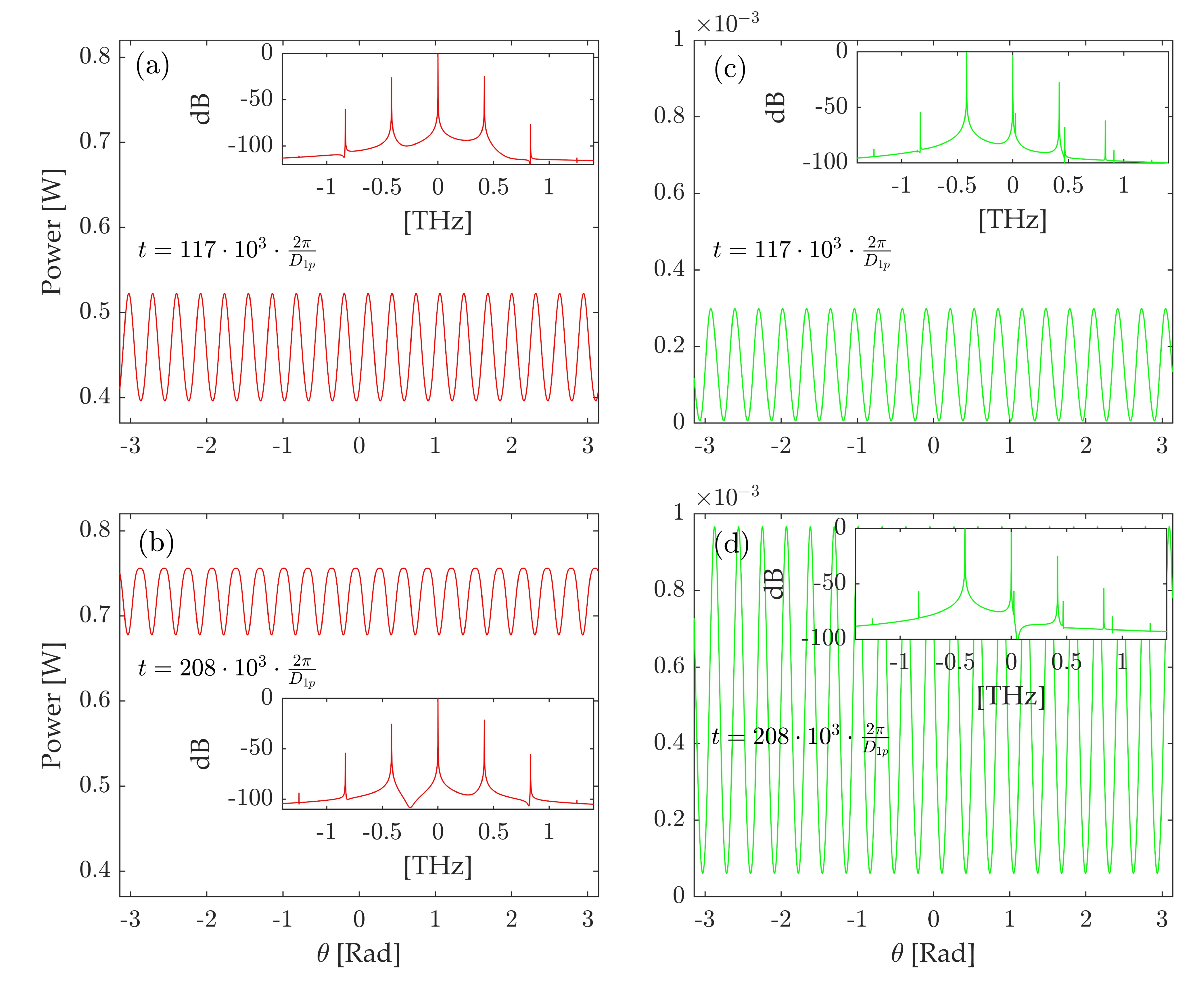}
		\caption{{\bf Quasi-stable Turing rolls and sparse combs.}
			Numerically computed real space intensity distributions over the micro-ring circumference of the intensities of the pump (red) and the second harmonic (green), and the corresponding spectra in the insets at two moments of time. Parameters are  the same as for Figs.\,4b, 4e in the main text.
			Note that $2\pi/D_\mathrm{1p}$ is the round trip time at the pump frequency. Amplitudes of the Turing roll patterns are breathing with time, but the pattern structure remain stable.
		}
		\label{fig1s}
	\end{figure*}  
	
	\newpage
	
	\begin{figure*}[h]
		\centering
		\includegraphics[width=0.9\textwidth]{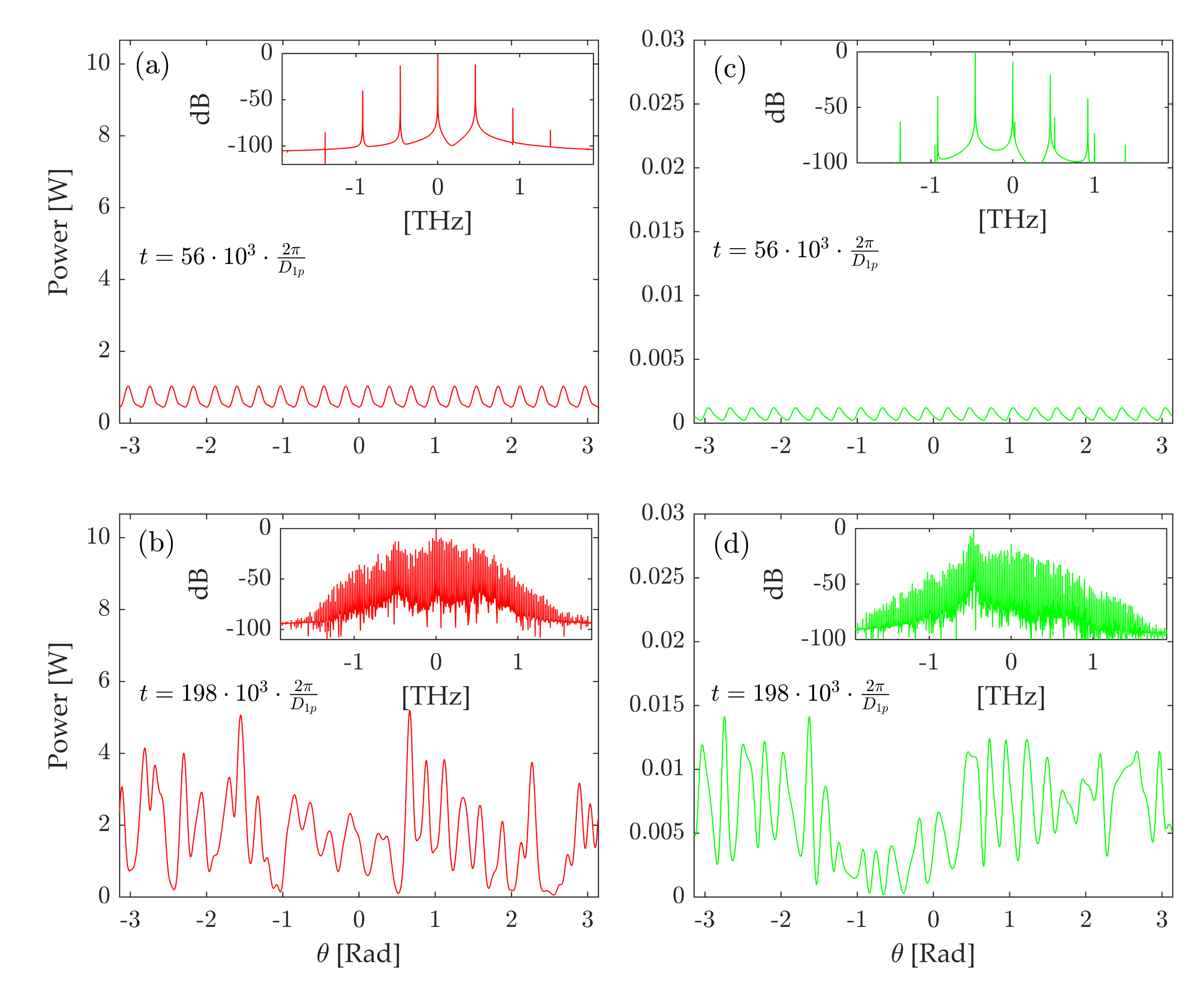}
		\caption{{\bf Instabilities of Turing rolls and quasi-continuous comb spectra.}
			Numerically computed real space intensity distributions over the micro-ring circumference of the intensities of the pump (red) and the second harmonic (green), and the corresponding spectra in the insets at two moments of time. Parameters are the same as for Figs.\,4c, 4f in the main text.
			Note that $2\pi/D_\mathrm{1p}$ is the round trip time at the pump frequency. Initially quasi-stable Turing roll are excited with corresponding sparse spectra. However, subsequently these patterns become unstable and break-up into irregular in space and time structures with dense quasi-continuous spectra.
		}
		\label{fig2s}
	\end{figure*}  
	
	\newpage
	\begin{figure*}[h]
		\centering
		\includegraphics[width=0.9\textwidth]{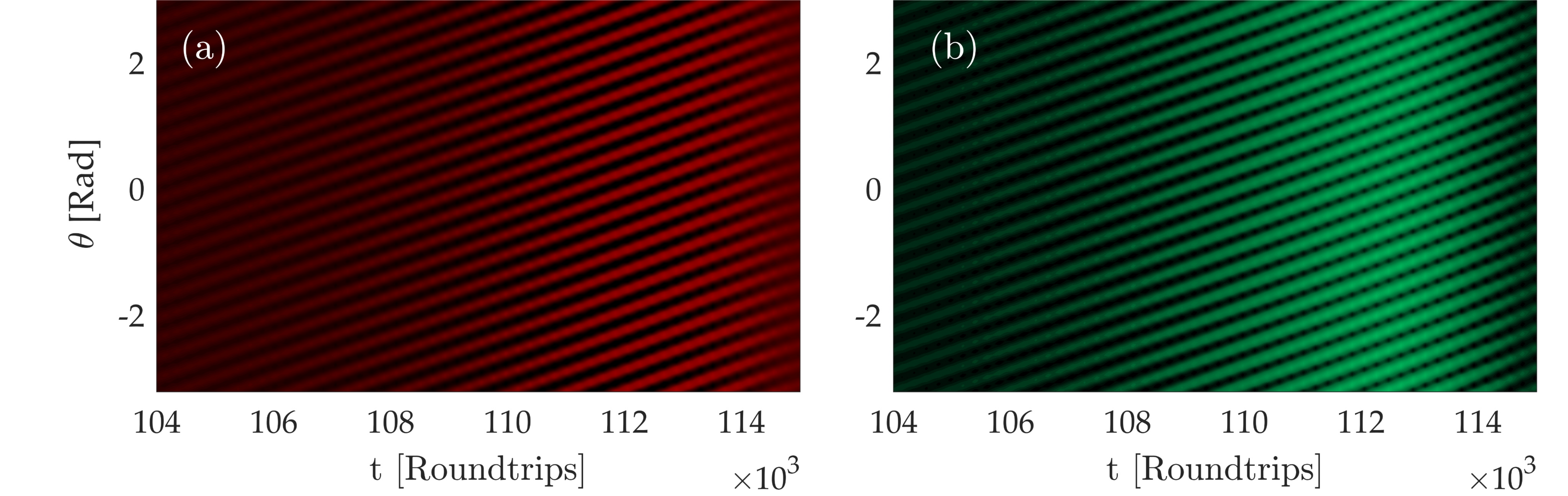}
		\caption{{\bf Estimating the locked FSR, FSR$_\mathrm{nl}$, from angular velocity of Turing rolls.} Quasi-stable Turing rolls corresponding to the data in Supplementary Fig.\,1 and in the main text Figs. 4b, 4e. (a)/(b) is the pump/2nd harmonic intensity. The structures shown were computed in the reference frame rotating with the $D_\mathrm{1p}$ frequency. Therefore the pattern tilt is the direct measure of the FSR$_\mathrm{p}$-FSR$_\mathrm{nl}$. After time $T=7000\times 2\pi/D_\mathrm{1p}\simeq 3.3654\times 10^{-7}$~s a maximum of the roll in both red and green travels over $\Phi\approx 2.2$~radians, so that FSR$_\mathrm{p}$-FSR$_\mathrm{nl}\simeq \frac{1}{2\pi}\frac{\Phi}{T}\simeq 1$~MHz.
		}
		\label{fig3s}
	\end{figure*}

	\begin{figure*}[h]
		\centering
		\includegraphics[width=0.9\textwidth]{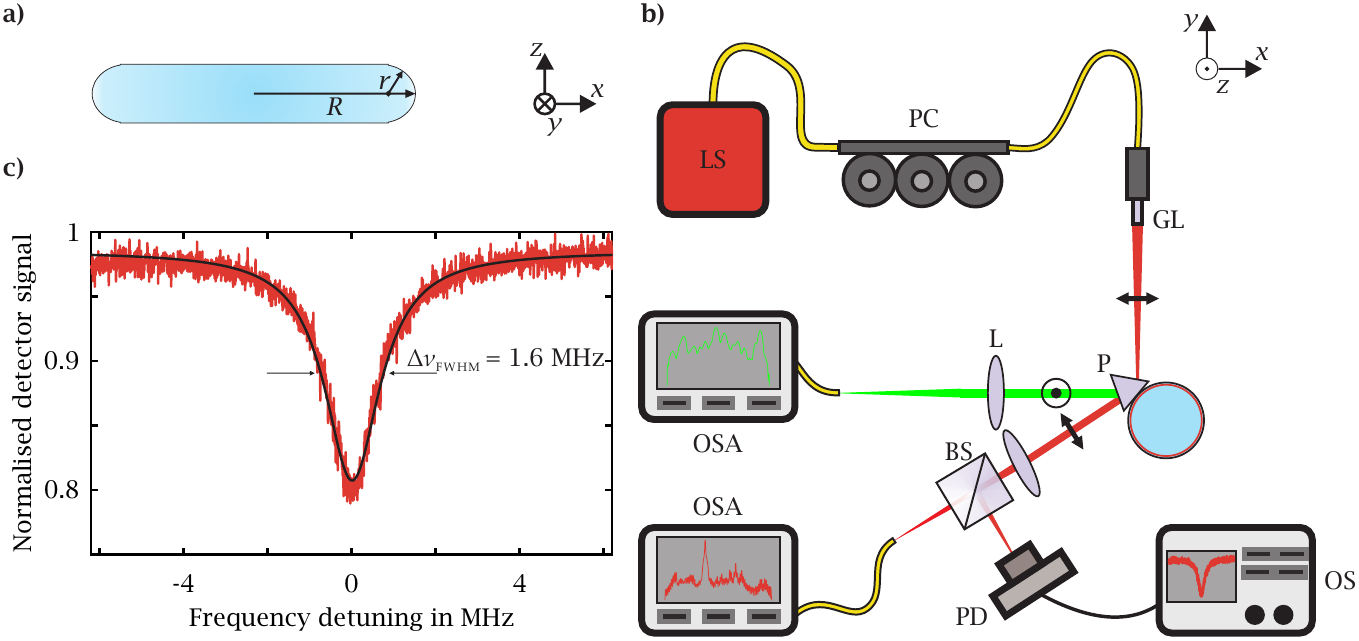}
		\caption{\textbf{Further experimental information.} \textbf{a)} Side-view of the microresonator geometry, which is characterised by the major radius $R$ and the minor radius $r$. \textbf{b)} Cavity resonance at very low input powers of a few $\upmu$W in critical coupling exhibiting a linewidth of 1.6~MHz and a maximum coupling efficiency of 18~\%. \textbf{c)} Experimental setup used for all the measurements in this contribution. Near-infrared light from a fiber-coupled laser source (LS) with polarisation controllers (PC) is focused through a gradient-index lens (GL) to couple light polarised in the x-y-plane to a microresonator by total internal reflection at the base of a rutile prism (P). When the resonator is heated to elevated temperatures, second-harmonic light is generated efficiently. Owing to the large birefringence of the prism, the outcoupled o-polarised light around the pump wavelength and the e-polarised light around the second-harmonic wavelength are separated spatially. The out-coupled light around the pump (red beam) then passes a beam splitter (BS): a small portion of it is focused on a photodetector (PD) connected to an oscilloscope (OS) to observe the transmission spectrum of the resonator. A larger portion enters an optical spectrum analyser (OSA) to observe the generated frequency spectrum. The light generated around the second-harmonic wavelength (green beam) is focused into another OSA using a lens (L) to be able to observe the generated spectrum in that wavelength region.
		}
		\label{fig4s}
	\end{figure*}  

\end{document}